\documentclass[a4paper,usenames,dvipsnames,11pt]{article}
\pdfoutput=1

\usepackage{jheppub}
\usepackage{slashed}
\usepackage{mathrsfs,booktabs,multirow,tabularx}
\usepackage{stmaryrd}
\usepackage{xspace}
\usepackage{fancyvrb}
\usepackage[makeroom]{cancel}
\renewcommand\arraystretch{1.1}

\def\beq{\begin{equation}}
\def\eeq{\end{equation}}
\def\beqn{\begin{eqnarray}}
\def\eeqn{\end{eqnarray}}
\def\abs#1{\left|#1\right|}

\newcommand\sss{\scriptscriptstyle}

\newcommand\as{\alpha_{\sss S}}

\newcommand\bQ{\bar{Q}}

\newcommand\aNLO{{\sc\small MadGraph5\_aMC@NLO}}
\newcommand\mf{{\sc\small MadFKS}}
\newcommand\ml{{\sc\small MadLoop}}
\newcommand\ct{{\sc\small CutTools}}
\newcommand\nin{{\sc\small Ninja}}
\newcommand\aMCSusHi{{\sc\small aMCSusHi}}
\newcommand\SusHi{{\sc\small SusHi}}

\newcommand\PYe{{\sc\small Pythia8}}
\newcommand\HWs{{\sc\small Herwig6}}
\newcommand\FJ{{\sc\small FastJet}}

\newcommand\pt{p_{\sss T}}
\newcommand\kt{k_{\sss T}}
\newcommand\Njet{N_{jet}}
\newcommand\FxFxM{{\rm FxFx}_{\sss\rm M}}
\newcommand\FxFxMT{{\rm FxFx}_{\sss\rm MT}}
\newcommand\FxFxEFT{{\rm FxFx}_{\sss\rm EFT}}
\newcommand\incM{{\rm inc}_{\sss\rm M}}
\newcommand\incMT{{\rm inc}_{\sss\rm MT}}
\newcommand\incEFT{{\rm inc}_{\sss\rm EFT}}
\newcommand\XFxFxM{$\FxFxM$}
\newcommand\XFxFxMT{$\FxFxMT$}
\newcommand\XFxFxEFT{$\FxFxEFT$}
\newcommand\XincM{$\incM$}

\newcommand\XincEFT{$\incEFT$}
\newcommand{\Qshow}{Q_{\text{sh}}}
\newcommand{\Qshowmax}{\Qshow}

\title{Heavy-quark mass effects in Higgs plus jets production}

\author[a]{R. Frederix,}
\author[b]{S. Frixione,}
\author[c]{E. Vryonidou,}
\author[d]{and M.  Wiesemann}
\affiliation[a]{Physik Department T31, Technische Universit\"at M\"unchen, 
James-Franck-Str. 1,\\ D-85748 Garching, Germany}
\affiliation[b]{INFN, Sezione di Genova, Via Dodecaneso 33, I-16146, 
Genoa, Italy}
\affiliation[c]{Centre for Cosmology, Particle Physics and Phenomenology 
(CP3),\\Universit\'e catholique de Louvain, B-1348 Louvain-la-Neuve, Belgium}
\affiliation[d]{Physik-Institut, Universit\"at Z\"urich, 
Winterthurerstrasse 190, 8057 Zurich, Switzerland}

\emailAdd{rikkert.frederix@tum.de}
\emailAdd{Stefano.Frixione@cern.ch}
\emailAdd{eleni.vryonidou@uclouvain.be}
\emailAdd{mariusw@physik.uzh.ch}

\abstract{We study the production of a Standard Model Higgs boson
in the gluon-fusion channel at the 13~TeV LHC. Our results are accurate
to the next-to-leading order in QCD, bar for the lack of some two-loop
amplitudes, for up to two extra jets and are 
matched to the \PYe\ Monte Carlo. We address the impact, at the level
of inclusive rates and of differential distributions, of the merging 
of samples characterised by different final-state multiplicities, and
of the effects induced by top and bottom masses through heavy-quark
loop diagrams. We find that both the merging and the heavy-quark masses
must be included in the calculation in order to realistically predict 
observables of experimental interest.
}


\preprint{
\begin{flushright}
CP3-16-11\\
MCNET-16-10\\
TUM-HEP-1040/16\\
ZU-TH-10/16\\
\today
\end{flushright}
}

\begin{document}
\maketitle
\flushbottom

\section{Introduction\label{sec:intro}}
The properties of the narrow-width particle discovered in 2012 by 
ATLAS~\cite{Aad:2012tfa} and CMS~\cite{Chatrchyan:2012xdj} at the LHC 
have so far been found to be fully compatible with those of the Higgs
boson in the Standard Model, where the main production mechanism proceeds 
through heavy-quark (denoted by $Q$ henceforth) loops in gluon fusion. 
The latter is an example of a so-called loop-induced process, in which 
{\em squared} one-loop amplitudes enter already at the Born level. The Higgs 
is attached to the loops by means of an $HQ\bQ$ vertex, hence the requirement 
that the quark be heavy, in order to have a sizable Yukawa coupling $y_Q$.
In the SM, contributions where $Q$ is the top quark are thus largely
dominant; however, those where bottom quarks circulate in the loops are 
not entirely negligible either, predominantly because they can interfere 
with the $Q=t$ ones. Incidentally, this mechanism is therefore responsible 
for a much more involved situation in BSM theories that feature a 
larger $y_b/y_t$ ratio w.r.t.~that of the SM.

Ignoring for the moment these issues, and considering only the 
$Q=t$ contributions, a very significant simplification may occur in 
the computation of the Higgs fully-inclusive cross section. The 
condition $m_h < 2m_t$ implies that the top degrees of freedom 
can be safely integrated out. By doing so, the three-point one-loop $ggH$ 
function is identified with an elementary vertex (and likewise for the four-
and five-point functions), whose corresponding interaction Lagrangian is 
added to that of the SM (minus the top) to define an effective 
field theory (EFT). From a more physical viewpoint, 
this is equivalent to saying that the virtuality of the $gg$ system,
equal to the Higgs mass, is not large enough to resolve the top-quark
loop, whose kinematical complications can therefore be ignored.  In this way, 
at any given order (in perturbative QCD) EFT computations feature one 
loop less than those in the full theory: for example, the Born $gg\to H$ 
process is associated with a tree-level (as opposed to one-loop) diagram. 
EFT results at the NLO in QCD have proved (somehow heuristically) to be an 
excellent approximation of those obtained in the SM~\cite{Djouadi:1991tka,
Dawson:1990zj,Spira:1995mt,Spira:1995rr}, once top-mass effects are 
accounted for at the LO. This has given a strong motivation 
to pursue calculations at yet higher orders, and predictions have thus been 
obtained at the  NNLO~\cite{Harlander:2002wh,Anastasiou:2002yz,
Ravindran:2003um} and even at the N$^3$LO~\cite{Anastasiou:2015ema}. 

While such theoretical results are extremely significant, it is important
to bear in mind that a deeper understanding of the $m_t\to\infty$ regime 
would be desirable; moreover, the knowledge of differential distributions is 
especially crucial. Firstly, these must be used in the computation of the 
acceptance corrections that allow one to relate the visible and the total 
cross sections. Secondly, they can be employed either in the direct comparison
to measured differential data (see e.g.~refs.~\cite{Aad:2015tna,Aad:2015lha,
Aad:2014tca,Aad:2014lwa,Khachatryan:2015yvw,Khachatryan:2015rxa} for
recent LHC results), which will become more numerous and precise in the near 
future, or in the context of multivariate analyses. Thirdly, and connected
to the previous item, experimental data may have to be characterised 
in terms of fixed jet multiplicities, and theory predictions must match 
such a characterisation. At the differential level, in those phase-space 
regions where a Higgs and/or jets with sizable energy can be found
(the rule of thumb being e.g.~$\pt(H)\gtrsim m_H$), the EFT is an
increasingly poor approximation, since the virtuality of the incoming-parton
pair becomes sufficiently large to resolve even top-quark loops. It is
therefore mandatory to obtain predictions for all the relevant observables
by retaining the exact heavy-quark mass dependences in the loops; in this way, 
one is also able to treat bottom-quark contributions in a sensible manner.

In the course of the past few years, full-SM predictions that lift the
EFT approximation have started to appear. For inclusive cross sections, 
the NNLO results of refs.~\cite{Marzani:2008az,Harlander:2009my,Pak:2009dg} 
(obtained by means of an asymptotic expansion in $1/m_t$) have confirmed 
earlier findings at the NLO, that top-mass effects have a very small impact. 
Differential 
observables are much more laborious to compute, and only a handful of results 
exist that go beyond the $m_t\to\infty$ approximation (conversely, there has 
been a significant activity within the EFT, too vast to be covered here; for 
recent reviews, see e.g.~refs.~\cite{Heinemeyer:2013tqa,Wiesemann:2015fxy,
Grazzini:2015jek}). At fixed NLO, subleading $1/m_t$ predictions are available 
for the Higgs and hardest-jet $\pt$~\cite{Harlander:2012hf,Neumann:2014nha}. 
Mass effects are also accounted for in analytically-resummed transverse
momentum spectra~\cite{Mantler:2012bj,Grazzini:2013mca,Banfi:2013eda}.
As far as perturbative computations matched to parton showers are concerned, 
multi-jet merged predictions at the leading order have been first presented 
in ref.~\cite{Alwall:2011cy} (see also ref.~\cite{Buschmann:2014sia}).
NLO+PS simulations have also been available for a while for inclusive $H+0j$
production, in the POWHEG~\cite{Bagnaschi:2011tu} and MC@NLO~\cite{MCatNLO408}
(see also ref.~\cite{Mantler:2015vba}) frameworks, with the latter 
subsequently employed~\cite{Buschmann:2014sia} within the 
MEPS@NLO~\cite{Hoeche:2012yf} merging scheme. Finally, NLO-accurate mass 
effects have been included in an NNLOPS implementation~\cite{Hamilton:2015nsa}.

The aim of this work is to assess the impact of heavy-quark masses on 
observables of current experimental interest, as well as to give realistic, 
hadron-level, differential predictions for both the Higgs boson and
several accompanying jets. This is achieved in the context of merged
simulations, where MC@NLO~\cite{Frixione:2002ik} samples of a given
$H+nj$ multiplicity are consistently combined by means of the FxFx
method~\cite{Frederix:2012ps} for $n=0,1,2$; the \PYe~\cite{Sjostrand:2007gs} 
parton shower is used throughout. Born and real-emission amplitudes
are computed exactly at one loop for all samples (therefore, the one-loop
amplitudes with the highest mutiplicity considered feature three final-state
partons, and enter the real corrections of the $n=2$ sample). Conversely,
the virtual amplitudes are computed exactly at two loops~\cite{Spira:1995rr,
Harlander:2005rq,Aglietti:2006tp} for $n=0$; when $n=1,2$, the one-loop EFT 
virtuals are used instead, rescaled by the ratio of the full over the EFT 
Born amplitude squared. While we note that our approach thus
employs all of the most accurate matrix elements available to date 
for $n\le 2$, it would be interesting to re-assess our
findings should the exact two-loop amplitudes with $n=1,2$ become
available\footnote{Such amplitudes {\em might} induce some differences,
typically in the kinematics regions characterised by large mass scales
(e.g.~above the top-quark threshold), w.r.t.~the approximated ones we 
have used, if effects that do not factorise the Born would turn out to be 
important, as was recently observed~\cite{Borowka:2016ehy} in a case that 
bears some analogy with the present one, namely di-Higgs hadroproduction.}.
In summary, the present work has a scope similar to that of 
refs.~\cite{Alwall:2011cy, Buschmann:2014sia}; it differs from the former in
that it features NLO effects, and from the latter in the merging scheme
employed, as well as in the facts that it includes (some of the) two-loop and
bottom-quark contributions, and that two-jet observables are also
NLO-accurate.

The paper is organised as follows. In sect.~\ref{sec:calc} we 
describe the nature of the simulations we have performed, as well
as their detailed setups, including input parameters. In sect.~\ref{sec:res}
we present predictions for inclusive rates and differential observables.
We draw our conclusions in sect.~\ref{sec:conc}.

\section{Details on the calculations\label{sec:calc}}
We consider four different types of simulations, all accurate to
NLO\footnote{Strictly speaking, in order to formally attain such an
accuracy in the full SM we would need to compute the virtual amplitudes 
at two loops, which we do only in the case of the lowest multiplicity.}
in QCD and matched to parton showers, which we list here roughly
in order of increasing complexity:
\begin{itemize}
\item inclusive $H+0j$ in the EFT\footnote{These are pure EFT results: 
there is no rescaling (e.g.~overall) by the full-SM over EFT Born 
matrix elements.}, denoted by {\bf inc}$_{\bf EFT}$;
\item inclusive $H+0j$ in the full SM, denoted by {\bf inc}$_{\bf M}$;
\item merged $H+nj$ ($n\le 2$) in the EFT, denoted by {\bf FxFx}$_{\bf EFT}$;
\item merged $H+nj$ ($n\le 2$) in the full SM, denoted by {\bf FxFx}$_{\bf M}$.
\end{itemize}
It is by now common practice~\cite{Grazzini:2013mca} in the context of 
matrix-element computations combined with resummation (be it analytical or
achieved through parton showers) to treat separately the contributions due to
squared top-loop amplitudes from those stemming from bottom loops (which thus
include the interference between top-loop and bottom-loop amplitudes).
This allows one to assign different resummation or shower scales to
such contributions, with the scale relevant to the latter (of the order
of a few bottom masses) smaller than that relevant to the former 
(of the order of $m_H$). One needs to keep in mind that this procedure
is pragmatic, but does not address directly the three-scale problem
posed by the bottom-loop contributions, a problem that must still 
be considered as open (for recent considerations, see 
e.g.~ref.~\cite{Melnikov:2016emg}). The presence of tagged final-state 
jets further complicates the picture, because it potentially introduces extra 
hard scales. This implies that the merging of samples stemming from bottom 
loops is necessarily quite heuristic, given the delicate role that mass 
scales play in merging procedures. In view of this, and of the fact 
that (even with a more sophisticated treatment of the three-scale problem
w.r.t.~what is done at present) the resummation or shower scale naturally
associated with bottom-loop contributions will result in relatively
soft jets, we have adopted the simple solution of treating these
contributions purely as an inclusive $H+0j$ sample\footnote{Note also
that, at variance with the case of top-induced contributions, the lack
of two-loop results for higher multiplicities is a serious problem in
the case of the bottom, since the rescaled-EFT one-loop matrix elements 
do not offer a viable approximation of the two-loop ones.}. Such a sample 
is then added to the samples stemming from top loops: to the inclusive one
to give \XincM, and to the FxFx-merged one to give
\XFxFxM. By adopting a rather sketchy notation, we can 
summarise the previous discussion as follows:
\beqn
\incM&=&(H+0j)_{t^2}+(H+0j)_{b^2}+(H+0j)_{bt}\,,
\label{incMdef}
\\
\FxFxM&=&{\rm FxFx}\Big[(H+0j)_{t^2}+(H+1j)_{t^2}+(H+2j)_{t^2}\Big]+
(H+0j)_{b^2}+(H+0j)_{bt}\,,\phantom{aaa}
\label{FxFxMdef}
\eeqn
with each sub-sample (in round brackets) meant to be NLO-accurate,
and the FxFx ``operator'' on the r.h.s.~of eq.~(\ref{FxFxMdef}) understood
to implement the FxFx merging. 
In a few occasions we shall also present full-SM results by retaining only 
top-loop contributions. By using the notation of eqs.~(\ref{incMdef}) 
and~(\ref{FxFxMdef}), these are therefore:
\beqn
\incMT&=&(H+0j)_{t^2}\,,
\label{incMTdef}
\\
\FxFxMT&=&{\rm FxFx}\Big[(H+0j)_{t^2}+(H+1j)_{t^2}+(H+2j)_{t^2}\Big].
\label{FxFxMTdef}
\eeqn

Our calculations are performed by means of \aNLO~\cite{Alwall:2014hca}, which
automates all ingredients relevant to the simulation of LO and NLO cross
sections, including the matching to parton showers.  The FKS
method~\cite{Frixione:1995ms,Frixione:1997np} (automated in the module
\mf~\cite{Frederix:2009yq}) for the subtraction of the infrared (IR)
singularities of real-emission matrix elements underpins all NLO-accurate
results. The computations of one-loop amplitudes are carried out by switching
dynamically between two integral-reduction techniques, 
OPP~\cite{Ossola:2006us} or Laurent-series expansion~\cite{Mastrolia:2012bu},
and TIR~\cite{Passarino:1978jh,Davydychev:1991va,Denner:2005nn}. 
These have been automated in the module \ml~\cite{Hirschi:2011pa}, which 
in turn exploits \ct~\cite{Ossola:2007ax} or \nin~\cite{Peraro:2014cba,
Hirschi:2016mdz}, together with an in-house implementation of the 
{\sc OpenLoops} optimisation~\cite{Cascioli:2011va}; we point out that, 
starting from v2.3.0, \ml\ can handle loop-induced processes 
automatically~\cite{Hirschi:2015iia}.

The status of \aNLO\ just described allows one to perform all simulations
outlined before without any human intervention in the EFT case, and with a
minimal amount of it when working in the full SM. However, the latter is 
{\em very} time consuming, owing to the presence of numerically-computed loop
amplitudes in all the contributions to, and the stages of, the computations,
which renders the generation of millions of events rather impractical on a
medium-size cluster. We have therefore proceeded as follows:
\begin{itemize}
\item All of the matrix elements relevant to the $(H+0j)_\alpha$ sub-samples, 
with $\alpha=t^2, b^2, bt$, have been replaced, by means of the \aMCSusHi\ 
script~\cite{Mantler:2015vba}, with their \SusHi~\cite{Harlander:2012pb}
analytically-computed counterparts. Note, in particular, that this
implies that two-loop virtuals are employed in such sub-samples,
which are integrated and unweighted directly.
\item The one-loop matrix elements for the other sub-samples, $(H+1j)_{t^2}$
and $(H+2j)_{t^2}$, are stored in a library generated by \ml. The library
includes a wrapper that allows one to call the individual matrix elements.
\item The sub-samples of the previous item are first integrated and
unweighted by working in the EFT. The resulting hard events are then 
reweighted by using the relevant\footnote{When very close to the IR limits
of the real-emission matrix elements, Born matrix elements are employed
in the reweighting, in order to ensure a more stable numerical behaviour,
and a marginally faster procedure. Furthermore, as is customary Borns 
are also used to reweight the virtuals.} full-SM over EFT ratios, with the
numerators computed on the fly using the \ml\ library constructed before.
This is the same procedure used to compute di-Higgs production in gluon
fusion in refs.~\cite{Frederix:2014hta,Maltoni:2014eza}.
\end{itemize}
A few comments are in order. Firstly, the reweighting procedure described
above does not entail any approximation. Its only implication is that the 
final event samples, given in input to the MC, are weighted rather than
unweighted (as it would be customary in \aNLO). Although a secondary unweighting
could be envisaged, we did not consider it in this work. Secondly, the
Higgs-plus-three-partons top-loop amplitudes enter (although in regions
that do not overlap kinematically) both the $(H+0j)_{t^2}$ contribution
(as real corrections) where they are computed by \SusHi, and the 
$(H+1j)_{t^2}$ contribution (as Borns) where they are computed by \ml.
Given the strict equivalence between \SusHi\ and \ml\ one-loop results, the 
$(H+1j)_{t^2}$ bit could also be computed with \SusHi; this would be faster, 
but rather more awkward from a procedural point of view, and we have therefore
refrained from doing it. Thirdly, the reweighting by the full-SM matrix
elements adopted in this paper is carried out by using, in part, software
routines which are not public. However, the hard events will be made 
publicly available\footnote{The interested reader is advised to
contact the authors.}. Furthermore, future public versions of \aNLO\
will contain NLO-accurate reweighting capabilities, equivalent to
the procedure adopted here.

\begin{table}[!h]
\begin{center}
\begin{tabular}{ll}
\toprule
$m_H=125$~GeV  & $m_t=172.5$~GeV  \\
$m_b=4.92$~GeV & PDFs: \texttt{NNPDF30\_nlo\_as\_0118}~\cite{Ball:2014uwa} \\
\bottomrule
\end{tabular}
\end{center}
\caption{\label{tab:input}
Input parameters for hard matrix elements. The pole top and bottom masses 
are used both in the propagators and in the Yukawas.}
\end{table}
The matrix elements have been computed by using the input 
parameters\footnote{Apart from the PDFs, these are in keeping with
the recommendations of the LHC Higgs Cross Section Working Group.
See {\tt https://twiki.cern.ch/twiki/bin/view/LHCPhysics/LHCHXSWG}.}
reported in table~\ref{tab:input}. The central renormalisation and 
factorisation scales ($\mu_0$) are set in merged simulations as dictated 
by the FxFx prescription~\cite{Frederix:2012ps}, and for inclusive results
equal to $H_{\sss T}/2$. In both cases, uncertainties 
are computed by varying independently the two scales in the range
\mbox{$[\mu_0/2,2\mu_0]$}; this is done automatically by \aNLO\ and
does not entail re-running, thanks to the method of
ref.~\cite{Frederix:2011ss}. The merging-scale dependence has been
assessed by comparing the results obtained by setting $\mu_Q=20$, $30$,
and $50$~GeV, with $30$~GeV being our default choice that corresponds
to the central predictions to be shown in sect.~\ref{sec:res}. We have
also considered the very extreme choice $\mu_Q=70$~GeV, but since the
corresponding results differ only marginally from those obtained 
with $\mu_Q=50$~GeV we refrain from reporting them here.

In \PYe, of which we have used both v8.210 and v8.212 (with 
identical results), the parameters have been set equal 
to their defaults, possibly as defined by the \aNLO\ interface. The Higgs 
boson is kept stable, and underlying events are not generated. We point out
that all the FxFx-related operations are now fully automated in \PYe;
more details can be found in ref.~\cite{Frederix:2015eii}, where
a phenomenology validation of the FxFx method is performed; at variance
with ref.~\cite{Frederix:2015eii}, no event oversampling is carried out
in the present paper. The jets that enter the definitions of physical
quantities are reconstructed at the hadron level by means of the
anti-$\kt$ algorithm~\cite{Cacciari:2008gp} with $R=0.4$ (as implemented
in \FJ~\cite{Cacciari:2011ma}), by keeping those with:
\beq
\pt(j)>30~{\rm GeV}\,,\;\;\;\;\;\;\;\;
\abs{\eta(j)}<4.4\,.
\label{jetcuts}
\eeq
We finally recall that \aNLO\ gives the user {\em some} control on the
event-by-event shower scale $\Qshowmax$ to be passed to the MC in the 
case of inclusive simulations, choosing it at random in the range
\beq
\alpha f_1\sqrt{s_0}\le\Qshowmax\le\alpha f_2\sqrt{s_0}\,,
\label{murange}
\eeq
where $\sqrt{s_0}$ is the Born-level partonic c.m.~energy, and $\alpha$,
$f_1$, and $f_2$ are user-defined numerical constants -- more details on
this can be found e.g.~in sect.~2.4.4 of ref.~\cite{Alwall:2014hca} and
in sect.~3.2 of ref.~\cite{Wiesemann:2014ioa}. This possibility is exploited
in the present context to assign different shower scales to the three
sub-samples that appear on the r.h.s.~of eq.~(\ref{incMdef}) (and thus
also to the two rightmost terms on the r.h.s.~of eq.~(\ref{FxFxMdef})),
in keeping with the procedure adopted in refs.~\cite{Harlander:2014uea,
Mantler:2015vba,Bagnaschi:2015bop,Bagnaschi:2015qta}. In particular, 
by choosing the settings suggested in ref.~\cite{Bagnaschi:2015bop}, 
the peak values of the shower-scale distributions for the three sub-samples 
are $\Qshowmax^{t^2}\sim m_H/2$, $\Qshowmax^{b^2}\sim 21$~GeV, and
$\Qshowmax^{bt}\sim 31$~GeV.

\section{Results\label{sec:res}}

\subsection{Inclusive rates\label{sec:rates}}
We start by presenting inclusive rates, which we collect
in table~\ref{tab:totFxFxb}. Each of the four leftmost columns presents
the results obtained with one of our standard merged or inclusive simulations.
In the rightmost column we report the contributions of the bottom-loop
induced processes to the full-SM merged and inclusive predictions.
In other words, for any given row the entries in the \XFxFxM\
and \XincM\ columns include the entry in the $\sigma_b$ column;
in turn, $\sigma_b$ is the cross section that emerges from the sub-samples
$(H+0j)_{b^2}$ and $(H+0j)_{bt}$  -- see eqs.~(\ref{incMdef}) 
and~(\ref{FxFxMdef}). All results are obtained with central hard and merging
scales; by using an error notation, we also show the fractional hard- and
merging-scale systematics, obtained as discussed in sect.~\ref{sec:calc}.

The first row of table~\ref{tab:totFxFxb} features the results obtained
without imposing any final-state cuts. The predictions in the next three 
rows are relevant to requiring a given number of exclusive ($\Njet=0,1$) 
or inclusive ($\Njet\ge 2$) jets among those that satisfy the cuts in 
eq.~(\ref{jetcuts}). The two rows at the bottom present cross sections 
in two phase-space regions defined by VBF-like cuts:
\beqn
&&{\rm VBF}_1:\;\;\;\;\;\;\;\;M(j_1,j_2)\ge 400~{\rm GeV}\,,
\;\;\;\;\Delta y(j_1,j_2)\ge 2.8\,,
\\
&&{\rm VBF}_2:\;\;\;\;\;\;\;\;M(j_1,j_2)\ge 600~{\rm GeV}\,,
\;\;\;\;\Delta y(j_1,j_2)\ge 4\,,
\eeqn
where $j_1$ and $j_2$ denote the hardest and second-hardest
jet of the event, respectively.

We observe that the hard-scale uncertainties of the merged results
are rather similar to those of the inclusive ones, if only slightly
larger, consistently with the fact that they receive contributions 
from matrix elements that feature higher powers of $\as$ w.r.t.~those
that enter the inclusive samples. Such uncertainties are significantly
larger, for all of the acceptance regions considered, than the merging-scale
ones, in spite of the fact that the latter are associated with a 
very conservative choice for the range of $\mu_Q$. This is particularly
striking in the case of the two VBF-like regions, for which the small
merging-scale dependence implies that the descriptions, given by the matrix 
elements and by \PYe, of kinematic configurations where jets tend to be 
close to the beam line are mutually quite consistent with each other.
This is clearly an MC-dependent statement, and is e.g.~at variance with
what has been found in the past with \HWs\ both at the LO and the
NLO~\cite{FFM}: it is wise to always assess the merging-scale
systematics associated with a given parton shower, especially in 
longitudinally-dominated regions.
\renewcommand\arraystretch{1.3}
\begin{table}[t]
\begin{center}
\resizebox{\columnwidth}{!}{%
\begin{tabular}{c|ccccc}
\toprule
 &  $\FxFxM$ & $\FxFxEFT$ & 
    $\incM$  & $\incEFT$  & $\sigma_b$ \\\midrule
Total 
    & $32.83^{+24.9\%}_{-19.5\%}{}^{+1.3\%}_{-2.6\%}$
    & $33.02^{+23.3\%}_{-18.8\%}{}^{+1.4\%}_{-2.4\%}$
    & $31.13^{+21.0\%}_{-18.2\%}{}$
    & $31.31^{+19.7\%}_{-17.6\%}{}$
    & $-2.05^{+2.9\%}_{-8.9\%}$
    \\
$\Njet=0$ 
    & $19.75^{+23.6\%}_{-18.7\%}{}^{+2.4\%}_{-0.5\%}$
    & $20.37^{+21.8\%}_{-18.0\%}{}^{+2.3\%}_{-0.3\%}$
    & $20.65^{+20.1\%}_{-18.0\%}$
    & $21.20^{+18.8\%}_{-17.3\%}$
    & $\;-1.97^{+5.7\%}_{-11.1\%}$
    \\
$\Njet=1$ 
    & $9.011^{+26.4\%}_{-20.5\%}{}^{+0.0\%}_{-5.8\%}$
    & $8.715^{+25.2\%}_{-19.9\%}{}^{+0.0\%}_{-6.1\%}$
    & $7.397^{+22.0\%}_{-18.6\%}$
    & $7.136^{+21.1\%}_{-18.0\%}$
    & $\!\!-0.10^{+27\%}_{-77\%}$
    \\
$\Njet\ge 2$ 
    & $4.061^{+30.4\%}_{-25.0\%}{}^{+0.0\%}_{-5.7\%}$
    & $3.935^{+29.7\%}_{-24.8\%}{}^{+0.0\%}_{-5.7\%}$
    & $3.083^{+31.9\%}_{-21.7\%}$
    & $2.972^{+32.1\%}_{-21.8\%}$
    & $0$
    \\
VBF$_1$ 
    & $0.512^{+29.6\%}_{-26.0\%}{}^{+0.0\%}_{-3.8\%}$
    & $0.518^{+29.8\%}_{-25.9\%}{}^{+0.0\%}_{-5.1\%}$
    & $0.411^{+32.7\%}_{-22.0\%}$
    & $0.402^{+32.7\%}_{-22.0\%}$
    & $0$
    \\
VBF$_2$ 
    & $0.214^{+29.0\%}_{-26.4\%}{}^{+0.0\%}_{-2.3\%}$
    & $0.221^{+30.5\%}_{-26.7\%}{}^{+0.4\%}_{-5.0\%}$
    & $0.191^{+32.5\%}_{-21.7\%}$
    & $0.184^{+32.3\%}_{-21.6\%}$
    & $0$
    \\
\bottomrule
\end{tabular}
}
\end{center}
\caption{\label{tab:totFxFxb}
Rates (in pb) for the FxFx-merged and inclusive simulations,
in both the full SM and the EFT. The contributions to the full-theory
predictions due to bottom-quark loops ($\sigma_b$) are also reported 
separately; a null entry indicates a result of the same order as the error 
on the corresponding $\FxFxM$ (or $\incM$) entry. The fractional hard-scale 
(left) and merging-scale (right, where relevant) uncertainties are given 
with an error notation.
}
\end{table}
\renewcommand\arraystretch{1.1}

The impact of the bottom-induced contributions, measured by the
ratio of $\sigma_b$ over \XFxFxM\ or \XincM,
is non-negligible in the regions dominated by small
jet multiplicities, but irrelevant elsewhere. This is a consequence
of the choices made for the treatment of such contributions
(see sect.~\ref{sec:calc}), and consistent with the idea that
bottom-loop amplitudes should describe the emission of mostly soft jets.
As expected, the opposite behaviour is associated with merging,
whose impact becomes more significant the larger the jet multiplicity.
Even in the fully-inclusive case (first row of table~\ref{tab:totFxFxb}),
merged rates are larger by about 5.5\% w.r.t.~the inclusive ones.
The cross section thus increases (we recall that FxFx is a non-unitary 
prescription), and this increase, as we shall see from the differential
results of sect.~\ref{sec:obs}, is essentially driven by the contributions
of the sub-samples $(H+1j)_{t^2}$ and $(H+2j)_{t^2}$.

\begin{table}[t]
\begin{center}
\begin{tabular}{c|ccc}
\toprule
 & $\Njet=0$ & $\Njet=1$ & $\Njet\ge 2$ 
 \\\midrule
$\FxFxM$
    & $0.602$
    & $0.274$
    & $0.124$
    \\
$\FxFxEFT$
    & $0.617$
    & $0.264$
    & $0.119$
    \\
$\incM$
    & $0.663$
    & $0.238$
    & $0.099$
    \\
$\incEFT$ 
    & $0.677$
    & $0.228$
    & $0.095$
    \\
\bottomrule
\end{tabular}
\end{center}
\caption{\label{tab:ratios1}
Ratios of cross sections with a given number of inclusive or exclusive
jets, over total cross sections. 
}
\end{table}
In order to assess in a more transparent manner the effects of merging, 
we report in table~\ref{tab:ratios1} the ratios of the cross sections for 
a given number of inclusive or exclusive jets, over the corresponding
fully-inclusive cross sections. Note that these results are not independent
from those of table~\ref{tab:totFxFxb}, but are obtained from the latter
(by using only the central values). For example, the first, second, and third
entry from the left in the first row of table~\ref{tab:ratios1} are evaluated
by computing the ratio of the entry in the second, third, and fourth row,
respectively, of the first column of table~\ref{tab:totFxFxb}, over the first
entry in the same column of that table. As can be seen from 
table~\ref{tab:ratios1}, the merging decreases the $\Njet=0$ cross
sections by a {\em relative} 10\%, while it increases the $\Njet=1$ and
$\Njet\ge 2$ ones by a relative 15\% and 25\%, respectively. These
fractions are essentially independent of whether heavy-quark mass
effects are included or not. However, masses do have an impact for
a given class of calculations (merged or inclusive); in particular,
they decrease the $\Njet=0$ rates (owing to bottom-mass effects, as we
shall discuss below), while they increase the $\Njet=1$ and $\Njet\ge 2$ 
ones. 

\begin{table}[h!]
\begin{center}
\begin{tabular}{c|ccccc}
\toprule
 &  $\FxFxM/\FxFxEFT$ & 
    $\incM/\incEFT$ 
 &  $\FxFxMT/\FxFxEFT$ & 
    $\incMT/\incEFT$ \\\midrule
Total 
    & $0.994$
    & $0.994$
    & $1.056$
    & $1.060$
    \\
$\Njet=0$ 
    & $0.970$
    & $0.974$
    & $1.067$
    & $1.067$
    \\
$\Njet=1$ 
    & $1.034$
    & $1.037$
    & $1.045$
    & $1.050$
    \\
$\Njet\ge 2$ 
    & $1.032$
    & $1.037$
    & $1.032$
    & $1.037$
    \\
VBF$_1$ 
    & $0.988$
    & $1.022$
    & $0.988$
    & $1.022$
    \\
VBF$_2$ 
    & $0.968$
    & $1.038$
    & $0.968$
    & $1.038$
    \\
\bottomrule
\end{tabular}
\end{center}
\caption{\label{tab:ratios2}
Ratios of full-SM over EFT cross sections.
}
\end{table}
A different way of looking at mass effects stems from the computation
of the ratios of full-SM over EFT cross sections, whose results we
show in table~\ref{tab:ratios2} for both the merged and the inclusive
simulations. In the two rightmost columns of the table, the SM predictions
are computed by excluding bottom-loop contributions (i.e.~the 
numerators correspond to eqs.~(\ref{incMTdef}) and~(\ref{FxFxMTdef})).
The most striking feature of table~\ref{tab:ratios2} is
that such effects factorise almost exactly in the case of all the rates 
except the VBF ones; in other words, they are independent of whether
merging is carried out or not, except for longitudinally-dominated
configurations. In absolute value, mass effects are rather small when
both top- and bottom-loops are taken into account: almost completely 
negligible for fully-inclusive cross sections\footnote{This is due to a 
cancellation between top- and bottom-loop effects, as can be inferred 
from table~\ref{tab:totFxFxb}.}, and at most $\pm 3\%$ for the other 
rates considered here. The effects are larger, for the
fully-inclusive and lowest-jet multiplicity cases, when only massive
top quarks are included in the computations.

\subsection{Differential distributions\label{sec:obs}}
\begin{figure}
\vskip -3.6truecm
  \includegraphics[width=1.0\linewidth]{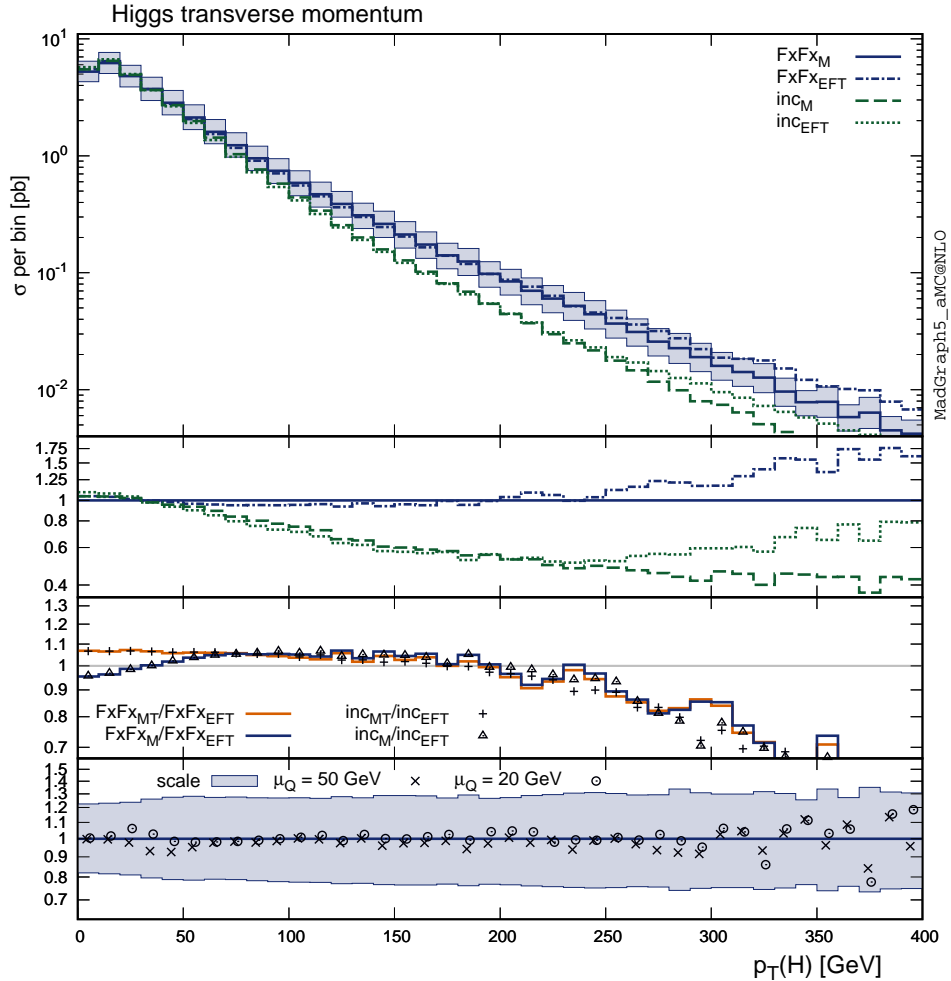}
\vskip -2.9truecm
  \caption{Inclusive Higgs transverse momentum.
The main frame displays the central results for our standard four 
predictions, as well as the hard-scale uncertainty band relevant to 
\XFxFxM. The upper inset presents ratios of the central
results over the \XFxFxM\ one. The middle insets highlights
heavy-quark mass effects in both merged and inclusive predictions.
The lower insets shows fractional hard- and merging-scale uncertainties
for \XFxFxM. See the text for further details.
}
  \label{fig:Higgsptinc}
\end{figure}

We now turn to presenting differential observables, constructed
with the Higgs and the jets four-momenta. We begin by considering
the transverse momentum of the Higgs boson; no cuts are imposed.
We display this quantity in fig.~\ref{fig:Higgsptinc}, whose layout is 
the following. In the main frame the central results of the four standard
simulations are shown: full-SM merged (blue solid histogram), EFT merged
(blue dot-dashed), full-SM inclusive (green dashed), and EFT inclusive 
(green dotted). In the case of \XFxFxM, the hard-scale uncertainty band
is also displayed as a shaded region. In the upper inset, the ratios 
\XFxFxEFT$/$\XFxFxM, \XincM$/$\XFxFxM, and \XincEFT$/$\XFxFxM\
are presented, by using the same patterns associated with the respective 
numerators in the main frame. The middle inset shows the ratios of the 
full-SM over EFT cross sections. Merged predictions are represented by
solid histograms: a blue (red) line is the result of using 
\XFxFxM\ (\XFxFxMT) as numerator. The analogues of these histograms
for the inclusive simulations are displayed as open triangles and 
crosses, respectively. The lower inset presents the fractional scale
uncertainties associated with \XFxFxM. Hard-scale systematics is
shown as a shaded region; as far as merging-scale dependence is concerned,
the ratios of the results obtained with $\mu_Q=20$~GeV and $\mu_Q=50$~GeV 
over the central $\mu_Q=30$~GeV prediction are given as open circles 
and crosses, respectively.

\begin{figure}[!ht]
\vskip -3.6truecm
  \includegraphics[width=1.0\linewidth]{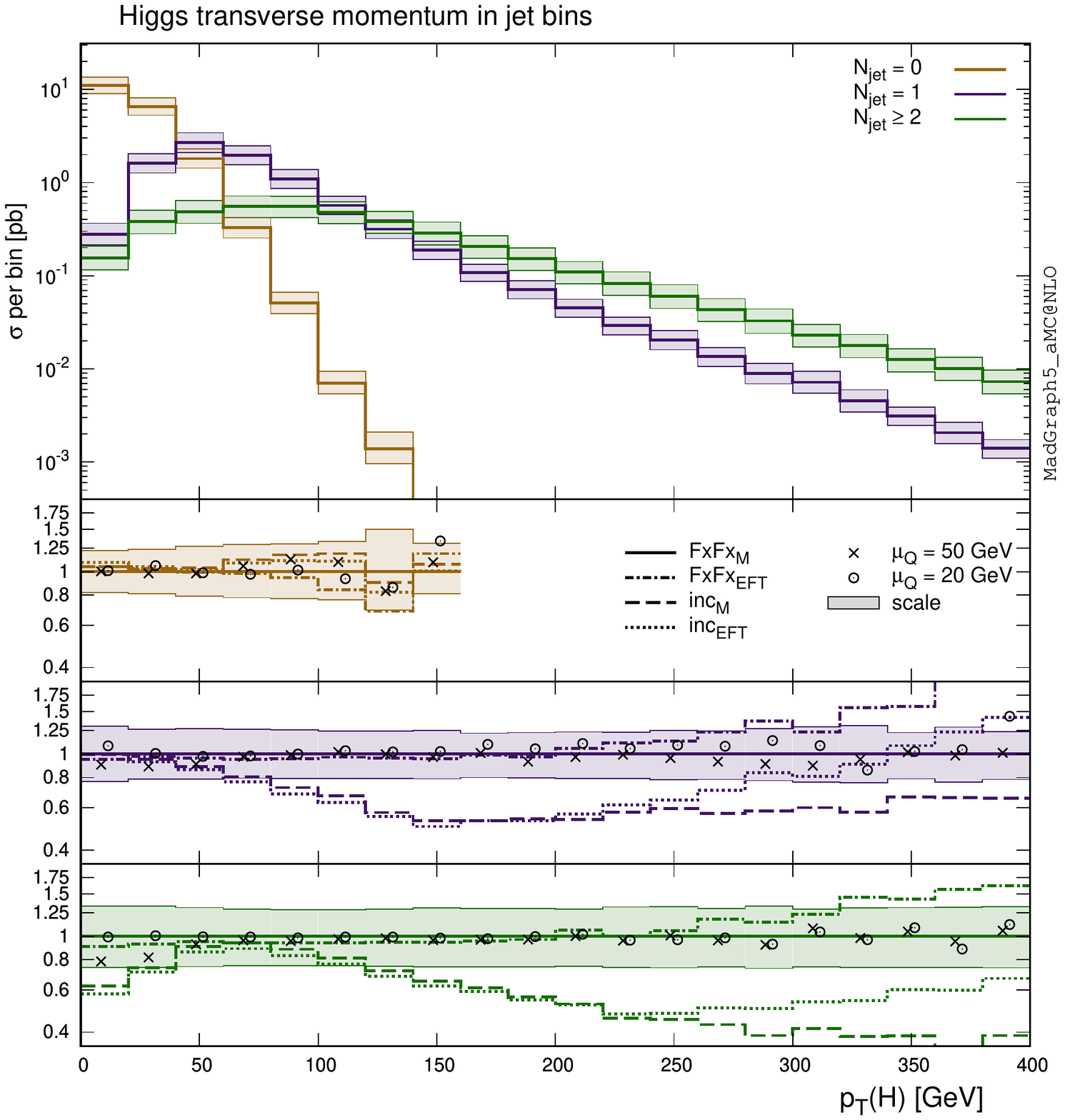}
\vskip -2.9truecm
  \caption{Higgs transverse momentum with $\Njet=0$, $\Njet=1$, 
and $\Njet\ge 2$. The main frame displays the central \XFxFxM\
predictions and their hard-scale uncertainty bands. Each inset
is relevant to one $\Njet$ multiplicity, and presents ratios computed
with the central \XFxFxM\ result as denominator. We show
the merged EFT and both inclusive predictions, and the fractional
hard- and merging-scale uncertainty of \XFxFxM.
See the text for further details.
}
  \label{fig:Higgsptjetbin}
\end{figure} 
The conclusions that can be drawn from fig.~\ref{fig:Higgsptinc} are
the following. The effects of multi-jet merging start to be quite
significant at $\pt(H)\gtrsim 50$~GeV; for $\pt(H)\gtrsim m_H$,
the inclusive results are outside of the hard-scale uncertainty band
of \XFxFxM. However, with increasing transerve momentum \XincEFT\ 
falls less rapidly than \XFxFxM, and as a consequence of that at 
about $\pt(H)\sim 350$~GeV it is again nearly compatible with the lower
border of the uncertainty band of the latter. This is ultimately
due to the fact that top-quark mass effects start to be visible
for $\pt(H)\gtrsim 250$~GeV, where they suppress the full-SM results
w.r.t.~their EFT counterparts. As can be seen from the middle inset,
by comparing the histograms with the symbols, heavy-quark mass effects
almost exactly factorise w.r.t.~the merging procedure: they affect
equally the merged and the inclusive predictions, which is quite
consistent with what has been already observed for inclusive rates
in sect.~\ref{sec:rates}. We note that this applies both to the large- 
and to the small-$\pt(H)$ region. In the latter, for $\pt(H)\lesssim 50$~GeV, 
the bottom-loop contributions do have a non-negligible impact on the shape 
of the distribution, in keeping with what previously 
found~\cite{Bagnaschi:2011tu,Mantler:2012bj,Grazzini:2013mca}.
Finally, the theoretical systematics that
affect the \XFxFxM\ result also have a similar pattern as those relevant
to inclusive rates: namely, on the whole transverse-momentum range considered,
hard-scale uncertainties largely dominate over merging-scale ones. The 
latter are in fact at most $\pm 10\%$, and typically much smaller than that.
In summary, multi-jet merging is more relevant to obtaining a sensible
prediction for the $\pt(H)$ spectrum than the exact treatment of heavy-quark
loops when statistics is limited, because this restricts one to relatively
small transverse momenta and/or to use bins of large widths. However,
as soon as one is able to access the high-$\pt$ region and to better
resolve the details of the spectrum (including its low-$\pt$ end),
then mass effects must mandatorily be taken into account.

The Higgs transverse momentum can also be observed in a more differential
manner, by requiring a given number of accompanying jets; this is 
potentially very relevant to experimental analyses which employ
$\Njet$ categorisation. We present our predictions for this quantity
in fig.~\ref{fig:Higgsptjetbin}, by requiring $\Njet=0$, $\Njet=1$,
and $\Njet\ge 2$. The layout of the figure is the following. In the
main frame, the solid histograms show the \XFxFxM\ results for the three
$\Njet$ cases, together with the respective hard-scale uncertainty bands. 
Each of the three insets, which have an indentical layout, is relevant to one 
$\Njet$ case. All results shown there are ratios of different numerators over 
the same denominator, that coincides with the central \XFxFxM\ prediction.
Dot-dashed, dashed, and dotted histograms are obtained by setting the
numerator equal to \XFxFxEFT, \XincM, and \XincEFT, respectively.
Open circles and crosses correspond to the $\mu_Q=20$~GeV and 
$\mu_Q=50$~GeV \XFxFxM\ cross sections, and thus give the fractional 
uncertainty associated with merging-scale choice. Finally, the shaded
area is the fractional hard-scale systematics.

Our findings can be summarised as follows. For $\pt(H)\gtrsim 40$~GeV,
the $\Njet=0$ result becomes smaller than the $\Njet=1$ one, and falls
very rapidly. At $\pt(H)\sim 100$~GeV, the $\Njet=1$ and $\Njet\ge 2$
predictions are comparable, with the latter becoming dominant as the
transverse momentum increases; the importance of a proper multi-jet
merging is therefore obvious. All types of simulation give very similar
results in the $\Njet=0$ bin; in particular, they are all compatible
with each other within the \XFxFxM\ hard-scale uncertainty (which
largely dominates over the merging-scale one). Things do change
when requiring non-null jet multiplicities. When $\Njet=1$, the
impact of merging is visible for $\pt(H)\gtrsim 50$~GeV, and that
of heavy-quark masses for $\pt(H)\gtrsim 250$~GeV. Although these
values are roughly the same as those relevant to the inclusive Higgs
$\pt$, the pattern of the corresponding effects is different; note,
in particular, that the \XincEFT\ prediction is significantly harder
than in the inclusive case. However, this feature should not be 
over-interpreted; in the presence of a tagged jet, \XincEFT\ will
depend to a certain extent on the MC adopted for the parton-shower phase.
As for $\Njet=0$, hard-scale systematics dominates.
The $\Njet\ge 2$ results follow roughly their $\Njet=1$ counterparts,
with two notable exceptions. Firstly, merging effects are present
also at small $\pt(H)$ (mostly owing to configurations where two jets
harder than the Higgs are present in the transverse plane, and recoil
against each other). Secondly, in that region the merging-scale dependence
is not completely negligible. We see, in particular, that the $\mu_Q=50$~GeV
result tends to be closer to the inclusive ones, which is consistent 
with the fact that such merging-scale choice emphasises more the role 
of the parton shower over that of the matrix elements, thus diminishing the
probability of finding two jets harder than the Higgs. 
Finally, we note that fairly small mass effects are also present at low 
$\pt(H)$ (although difficult to see in fig.~\ref{fig:Higgsptjetbin}), 
especially in the $\Njet=0$ and $\Njet\ge 2$ bins, where they are 
predominantly induced by bottom- and top-loop contributions, respectively.

\begin{figure}[!ht]
\vskip -3.6truecm
  \includegraphics[width=1.0\linewidth]{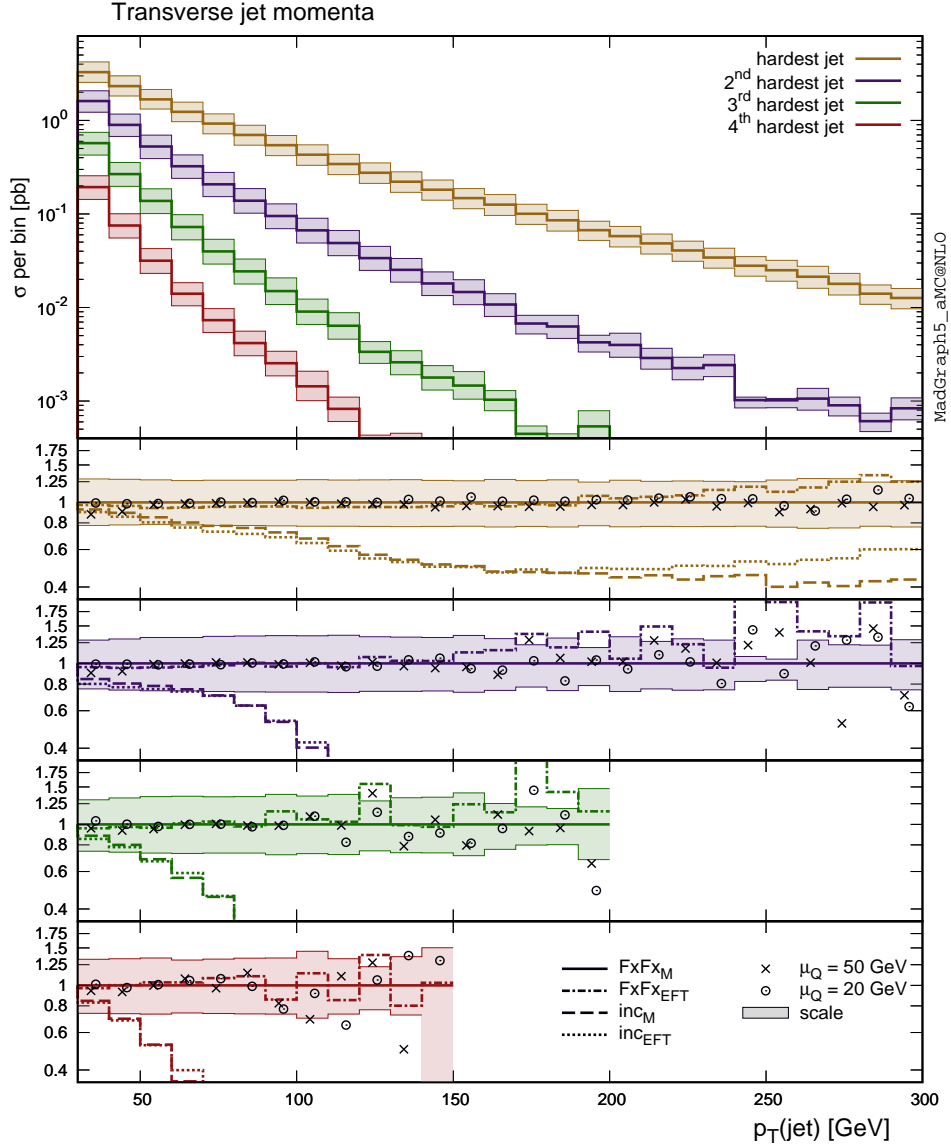}
\vskip -1.0truecm
  \caption{Same as in fig.~\ref{fig:Higgsptjetbin},
for the transverse momenta of the four hardest jets.
}
  \label{fig:ptjet1-4inc}
\end{figure} 
We next consider the transverse momenta of the four hardest jets, 
which we display in fig.~\ref{fig:ptjet1-4inc}; note that this figure 
has the same layout as fig.~\ref{fig:Higgsptjetbin}. These observables
are correlated, to different extents, to the inclusive or jet-binned Higgs 
transverse momentum previously discussed, and they thus have some
characteristics whose patterns are similar to those of the latter.
In particular, the onset of merging or of heavy-quark mass effects occurs
at transverse momenta of the same order as those relevant to the Higgs
spectra. On the other hand, the impact of merging is generally larger
than for the Higgs $\pt$ (except perhaps in the case of the hardest jet),
and it dramatically increases when one considers jets that are increasingly
subleading. This applies also to jets which are formally beyond the 
matrix-element accuracy of the simulations performed in this work,
and it need not be surprising. The same feature has been discussed
e.g.~in ref.~\cite{Frederix:2015eii} in the context of $Z/W\!+\!{\rm jets}$
production (which is analogous to Higgs production as far as merging is
concerned), and validated against data there. Essentially, the benefits of
merging extend beyond those naively expected from the underlying matrix
elements, because the latter provide the parton showers with much more
sensible initial conditions than those available in inclusive simulations.
In support of this fact, we also observe that the merging-scale dependence
is again much smaller than the hard-scale one, and all jets feature roughly
the same stability against $\mu_Q$ variations (the oscillations in some of
the $\pt$ tails are clearly not resulting from a pattern associated with
$\mu_Q$, but merely reflect a lack of statistics).

\begin{figure}[!ht]
\vskip -3.6truecm
  \includegraphics[width=1.0\linewidth]{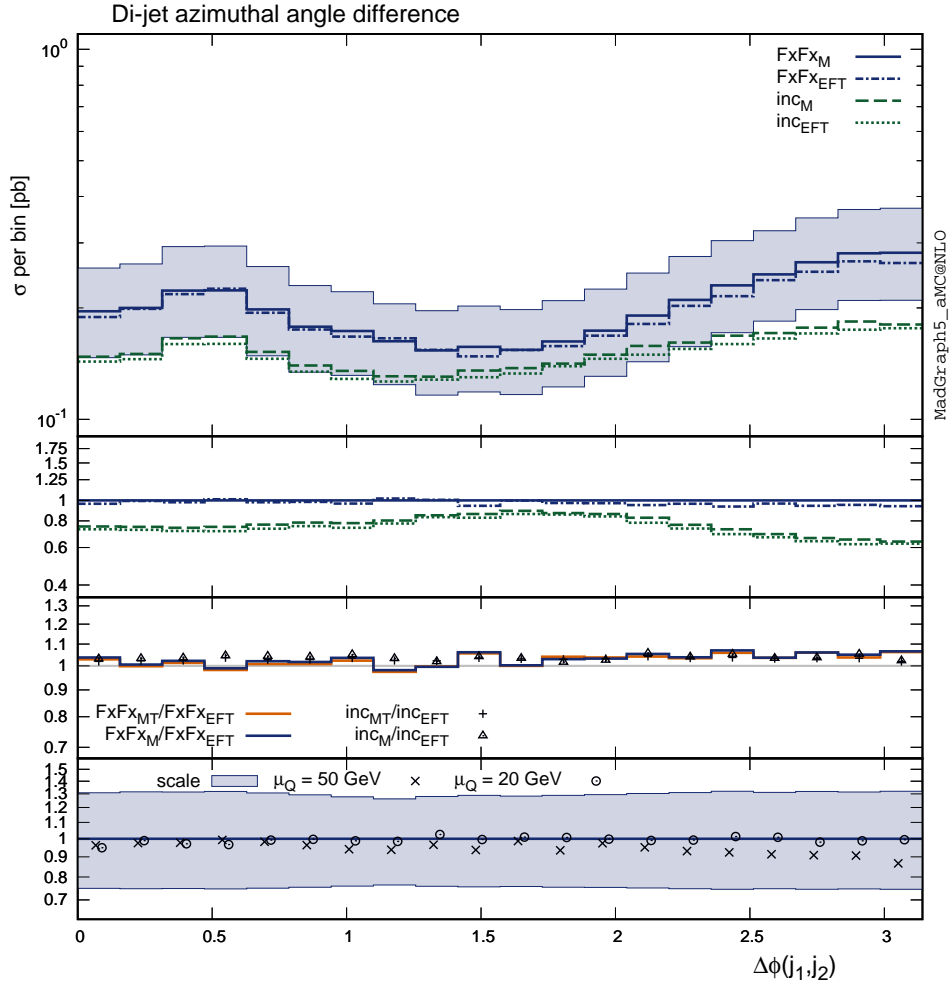}
\vskip -2.9truecm
  \caption{Same as in fig.~\ref{fig:Higgsptinc},
for the azimuthal distance between the two hardest jets.
}
  \label{fig:Deltaphij1j2}
\end{figure}
\begin{figure}[!ht]
\vskip -3.6truecm
  \includegraphics[width=1.0\linewidth]{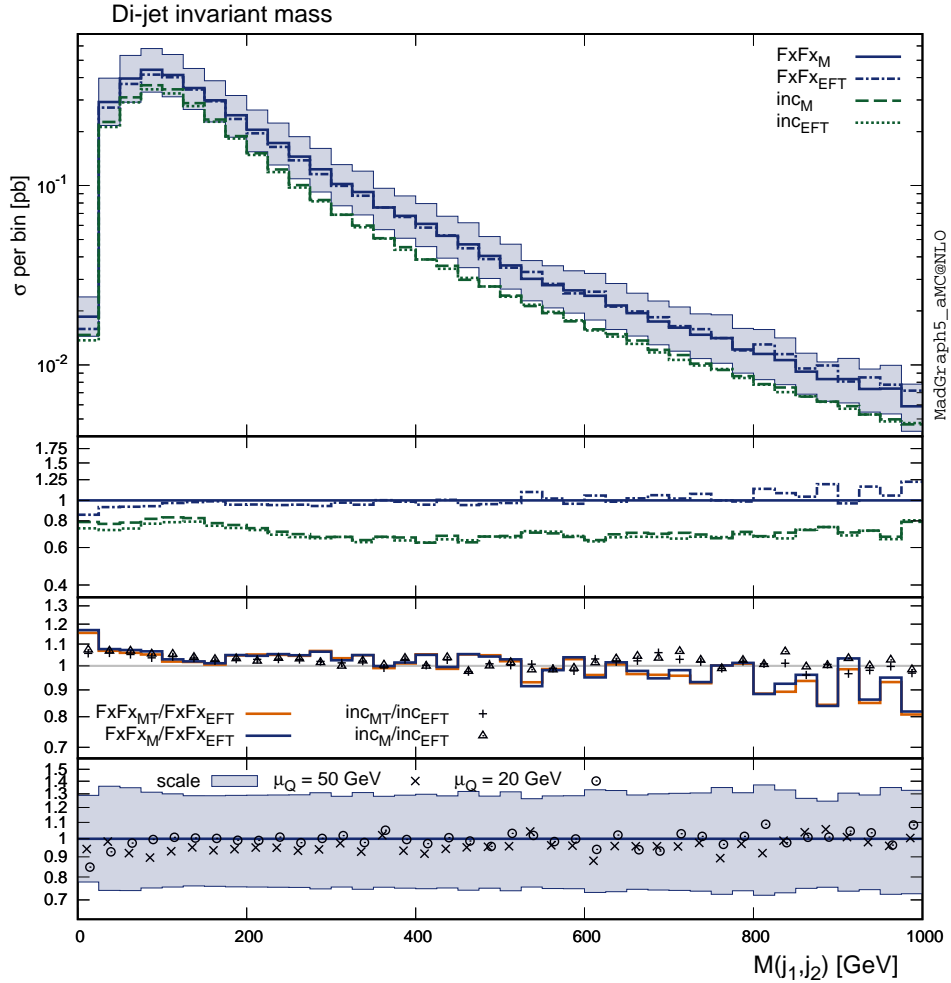}
\vskip -2.9truecm
  \caption{Same as in fig.~\ref{fig:Higgsptinc},
for the invariant mass of the pair of the two hardest jets.
}
  \label{fig:mjj}
\end{figure}

\begin{figure}[!ht]
\vskip -3.6truecm
  \includegraphics[width=1.0\linewidth]{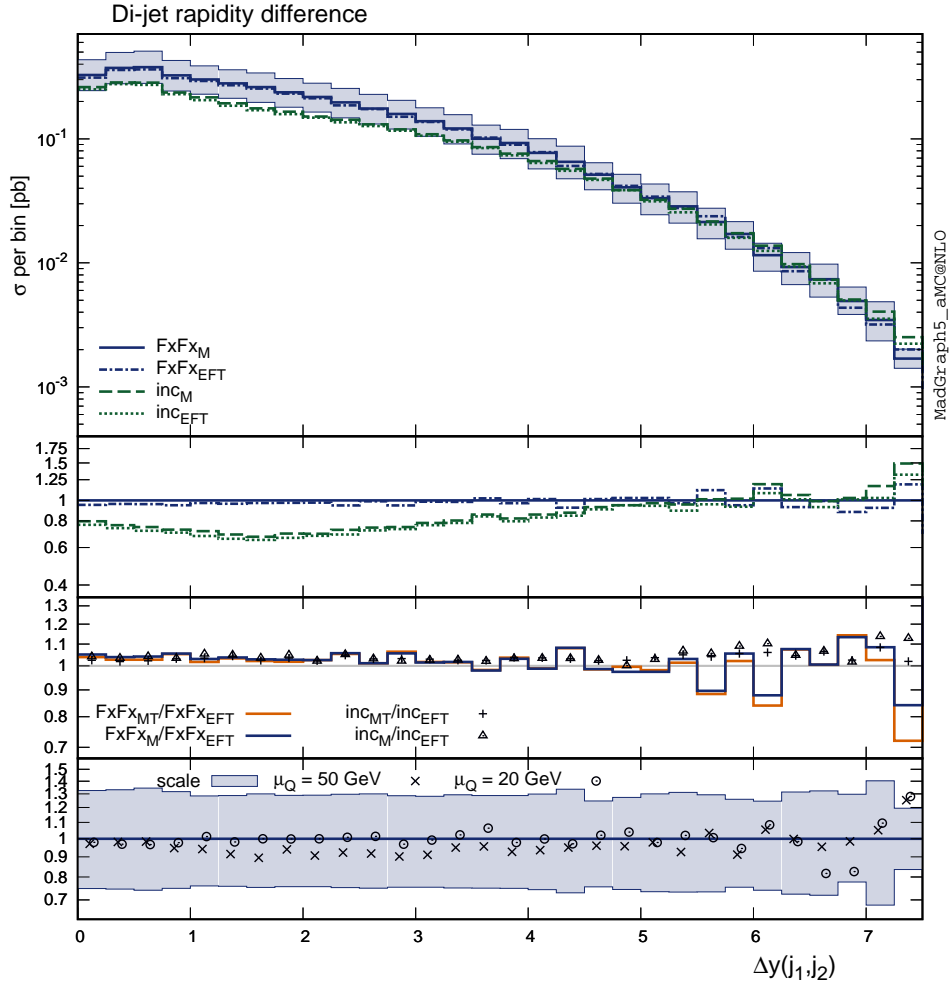}
\vskip -2.9truecm
  \caption{Same as in fig.~\ref{fig:Higgsptinc},
for the rapidity distance between the two hardest jets.
}
  \label{fig:Deltayj1j2}
\end{figure}

\begin{figure}[!ht]
\vskip -3.6truecm
  \includegraphics[width=1.0\linewidth]{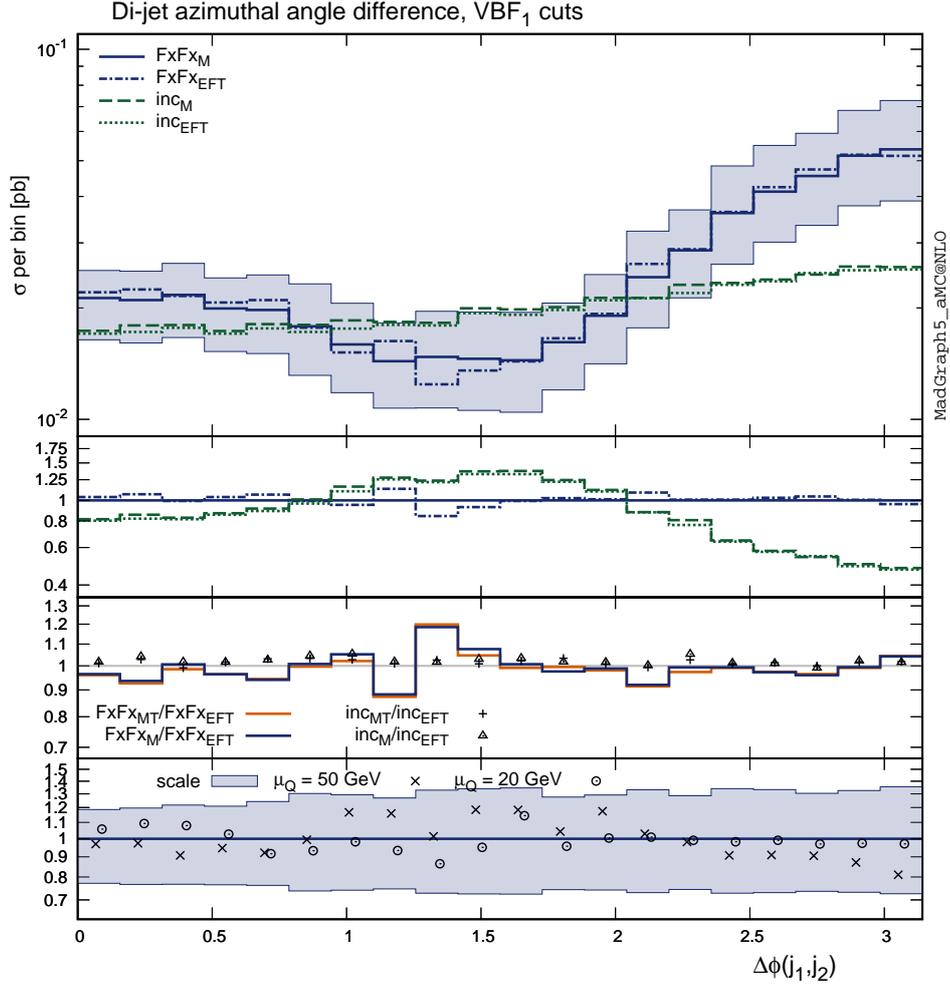}
\vskip -2.9truecm
  \caption{Same as in fig.~\ref{fig:Higgsptinc},
for the azimuthal distance between the two hardest jets,
within the VBF$_1$ acceptance region.
}
  \label{fig:Deltaphij1j2vbf1}
\end{figure}

 
We now move to consider a few correlations between the two 
hardest jets  of the event, which we present in 
figs.~\ref{fig:Deltaphij1j2}--\ref{fig:Deltaphij1j2vbf1}. All these
figures have exactly the same layout as fig.~\ref{fig:Higgsptinc}, 
to which we refer the reader for detailed explanations. We begin
in fig.~\ref{fig:Deltaphij1j2} with the azimuthal distance; this
observable has a distinctive shape, and can be used e.g.~to help assess
the presence of a BSM signal in data. As is clear from the figure,
the impact of merging is quite significant, especially shape-wise.
Conversely, heavy-quark mass effects are barely visible in the
central predictions, and amply within the hard-scale systematics.
This has to be expected, since the present observable is dominated
by configurations in which jets are produced at their $\pt$ thresholds.
As in all previous cases, merging-scale uncertainties are much smaller
than their hard-scale counterparts. Having said that, interestingly 
there is a trend at \mbox{$\Delta\phi(j_1,j_2)\to\pi$}, where the 
$\mu_Q=50$~GeV result is less peaked than the others. In other words,
for such a merging scale the two hardest jets are less back-to-back than
for smaller merging scales; this is the same phenomenon responsible for the
small-$\pt(H)$ behaviour when $\Njet\ge 2$, previously discussed
in fig.~\ref{fig:Higgsptjetbin}.

The invariant mass of the pair of the two hardest jets and their rapidity 
distance are shown in fig.~\ref{fig:mjj} and fig.~\ref{fig:Deltayj1j2},
respectively. 
These two observables are important in that they are used to 
define VBF regions; therefore, their shapes control the extent
to which a possible VBF signal is ``contaminated'' by that due
to gluon fusion. Apart from the obvious, observable-specific
differences, what has been said about \mbox{$\Delta\phi(j_1,j_2)$}
applies to \mbox{$M(j_1,j_2)$} and \mbox{$\Delta y(j_1,j_2)$} as well.
In particular, the effects of merging are evident, and especially
prominent in the case of the rapidity correlation. They
harden the spectrum of the invariant mass distribution, most notably
at intermediate masses (\mbox{$\sim 150-400$~GeV}), and decrease
the rapidity separation between the two jets, so that the merged
results have relatively more events than the inclusive ones towards
small \mbox{$\Delta y(j_1,j_2)$}. Heavy-quark mass effects are only
visible in the invariant-mass spectrum (see the middle inset of 
fig.~\ref{fig:mjj}), where size-wise they are roughly as important 
as the merging ones. They are responsible for a softening of the
distribution w.r.t.~the EFT behaviour, which softening in the large-mass
tail seems to affect only the merged predictions. While lack of
statistics may be an issue in this region, this trend is related
to the breakdown of the factorisation of mass effects remarked in 
sect.~\ref{sec:rates} for the case of VBF-like inclusive cross sections.
This is most likely due to an increased difference between the
high-multiplicity matrix elements computed with or without mass effects, 
since in this kinematic region multiple large scales become relevant 
in the computation of the loop amplitudes.
Finally, also for the two observables discussed here the merging-scale
systematics is much smaller than that due to hard-scale variations.
We remind the reader that, in the case of \mbox{$\Delta y(j_1,j_2)$},
this is a statement that depends rather strongly on the MC adopted; 
see sect.~\ref{sec:rates} for related observations on VBF inclusive rates.

We finally present in fig.~\ref{fig:Deltaphij1j2vbf1} the azimuthal 
distance between the two hardest jets within VBF$_1$ cuts.
We find patterns which are similar to those observed in the case
of its inclusive companion of fig.~\ref{fig:Deltaphij1j2}. However,
by comparing in particular the upper insets of the two figures, one
sees that in the present case the effects of merging are much larger
in absolute value. The almost complete de-correlation between the
two jets resulting from the inclusive simulations demands that merging
be carried out, in order to achieve physically-sensible results.
Heavy-quark mass effects are unimportant, and hard-scale uncertainty
is much larger than the merging-scale one, in spite of the latter
being more significant than in all the other cases considered so far.

\section{Conclusions\label{sec:conc}}
In this paper we have studied the production of a Standard Model 
Higgs boson in association with jets through the gluon-fusion channel,
presenting results for both inclusive rates and differential observables 
relevant to the 13~TeV LHC. We have done so by considering several types 
of simulations, in all of which matrix elements are matched to parton 
showers at the NLO accuracy in QCD according to the MC@NLO formalism.
We have systematically compared predictions that stem from inclusive 
samples with those based on the consistent merging, by means of the 
FxFx method, of sub-samples characterised by different parton-level
multiplicities, for up to two extra jets at the NLO. Within each type
of approach, inclusive or merged, we have evaluated the underlying
matrix elements both in the EFT where the Higgs couples directly 
to gluons, and in the SM by computing exactly the relevant top-
and bottom-loop amplitudes (except for the two-loop virtual contributions
to the one- and two-jet cross sections, which we have approximated).
All of our computations are carried out in the \aNLO\ framework,
and we have adopted the \PYe\ parton shower.

The primary conclusion of this work is that both merging and 
heavy-quark-mass effects must be taken into proper account in order
to predict realistically observables of experimental interest, that
will further our understanding of the particle discovered by ATLAS
and CMS. The impact of merging is especially prominent, while in
order to be sensitive to quark masses it is important to be able
to probe kinematic regions dominated by large-$\pt$ emissions,
and to have a good resolution power to allow for a sufficiently
fine binning, which helps uncover such effects at small transverse
momenta as well. Both features are characteristic of datasets with
substantial statistics.

For a sufficient number of events, merging and mass effects
provide one with very distinctive features, that have the further
advantage of being fairly observable-dependent, thus allowing for
stringent comparisons between predictions and measurements. The
theoretical systematics are dominated by the dependences on the
factorisation and renormalisation scales. Conversely, variations
of the FxFx merging scale, which we have carried out in a very large
range in order to be conservative, are almost always negligible,
with a very few exceptions in which they are still smaller than 
those induced by the hard scales. However, we point out that this
is an observable- and MC-dependent conclusion; the assessement
of the merging-scale systematics is an important task that must 
always be performed.

\section*{Acknowledgements}
We are grateful to Valentin Hirschi for his help with the generation of the
\ml\ process library, and to Fabio Maltoni for his comments on the
manuscript. The work of RF is supported by the Alexander von
Humboldt Foundation, in the framework of the Sofja Kovaleskaja Award Project
``Event Simulation for the Large Hadron Collider at High Precision'', endowed
by the German Federal Ministry of Education and Research.  SF thanks the CERN
TH Division for hospitality during the course of this work.  The work of EV is
supported in part by the European Union as part of the FP7 Marie Curie Initial
Training Network MCnetITN (PITN-GA-2012-315877).  The work of MW is supported
in part by the Swiss National Science Foundation (SNF) under contract
200021-156585.  This work is supported in part by the ERC grant 291377
``LHCtheory: Theoretical predictions and analyses of LHC physics: advancing
the precision frontier''.

\phantomsection
\addcontentsline{toc}{section}{References}
\bibliographystyle{JHEP}
\bibliography{Hrefs.bib}

\end{document}